\documentclass{article}
\usepackage{spconf, amsmath, amsfonts, graphicx, float}
\usepackage{setspace}
\usepackage{breqn}
\usepackage[utf8]{inputenc}
\usepackage{amsmath}
\usepackage{amssymb}
\usepackage{xspace}
\usepackage{subfigure}
\usepackage{algorithm}
\usepackage{algpseudocode}
\usepackage{tabularx}
\usepackage{multirow}
\usepackage{booktabs}
\usepackage{hyperref}
\usepackage{footmisc}
\usepackage[misc]{ifsym}
\usepackage{makecell}
\usepackage{color}
\usepackage{cite}
\usepackage{enumitem}
\usepackage{etoolbox}
\patchcmd{\thebibliography}
  {\settowidth}
  {\setlength{\itemsep}{0pt plus 0.1pt}\settowidth}
  {}{}
\apptocmd{\thebibliography}
  {\small}
  {}{}

\usepackage{xcolor}

\definecolor{YenjuColor}{rgb}{1.0, 0.0, 0.0}
\definecolor{ShinjiColor}{rgb}{0.0, 0.0, 0.8}
\definecolor{AlexColor}{rgb}{0.8, 0.8, 0.0}

\usepackage{fancyhdr}
\makeatletter
\DeclareRobustCommand\onedot{\futurelet\@let@token\@onedot}
\def\@onedot{\ifx\@let@token.\else.\null\fi\xspace}

\makeatother

\title{SRTNET: TIME DOMAIN SPEECH ENHANCEMENT VIA STOCHASTIC REFINEMENT}
\name{Zhibin Qiu$^{1\star}$, Mengfan Fu$^{1\star}$, Yinfeng Yu$^{1,2}$, LiLi Yin$^{1}$, Fuchun Sun$^{2}$, Hao Huang$^{1\dagger}$\thanks{$^{\star}$ Equal contributions,
$^{\dagger}$ correspondence author.}}
\address{$^1$School of Information Science and Engineering, Xinjiang University, China \\  $^2$Department of Computer Science and Technology, Tsinghua University, China
}
\begin{document}
\ninept
\thispagestyle{fancy}
\maketitle
\setlength{\abovedisplayskip}{2.5pt}
\setlength{\belowdisplayskip}{2.5pt}
\begin{abstract}
Diffusion model, as a new generative model which is very popular in image generation and audio synthesis, is rarely used in speech enhancement. In this paper, we use the diffusion model as a module for stochastic refinement. We propose SRTNet, a novel method for speech enhancement via \textbf{S}tochastic \textbf{R}efinement in complete \textbf{T}ime domain. Specifically, we design a joint network consisting of a deterministic module and a stochastic module, which makes up the ``enhance-and-refine” paradigm. We theoretically demonstrate the feasibility of our method and experimentally prove that our method achieves faster training, faster sampling and higher quality.
Our code and enhanced samples are available at
https://github.com/zhibinQiu/SRTNet.git
\end{abstract}
\section{Introduction}
\label{sec:intro}
Speech enhancement (SE) using generative models has great potential.
Typical generative methods for SE are GAN-based methods~\cite{SEGAN,GAN1-SEAME,GAN2-SEAME,GAN3-SEAME,SERGAN,SAGAN,CP-GAN}, flow-based methods~\cite{Flow-Based-SEAME,Flow-Based1-SEAME} and VAE-based methods~\cite{VAE-SEAME,VAE1-SEAME,VAE2-SEAME,VAE3-SEAME,VAE4-SEAME,VAE5-SEAME}.
Diffusion model is another generative model which is very popular recently~\cite{chen2020wavegrad,kong2020diffwave}. However, it is rarely used for speech enhancement. The SE methods 
in time domain not only avoid the distortions caused by inaccurate phase information~\cite{phase-importance}, but also avoid the extra overhead of computing the T-F representation.~\cite{CD-SEAME} proposed a method for SE 
in time domain using the diffusion model, but it relies on the speech spectrogram of noisy, so it is not a complete time-domain method.
Moreover, in~\cite{CD-SEAME}, the diffusion model directly estimates the distribution of clean speech by optimizing the evidence lower bound (ELBO), which puts a large computational pressure on the diffusion model and leads to a lot of time consumption during the training phase.~\cite{Deblurring}, a method of image deblurring, proposed the predict-and-refine approach to reduce the computational pressure of the diffusion model while guaranteeing the quality of the image generation. 
Inspired by this, a joint network paradigm is designed, namely \textbf{``enhance-and-refine"} which is comprised of two sub-modules, deterministic module and stochastic module, respectively. The two modules are connected by the residual structure, and the noisy speech is initially enhanced after passing through the deterministic module. Afterward, the initial enhanced result passes through the residual structure and into the stochastic module for detailed refinement. We refer to the network consisting of these two modules as SRTNet.
SRTNet allows the diffusion model to act as a stochastic module to learn the residual distribution instead of the distribution of data directly, which significantly reduces the computational overhead of the diffusion model. In addition, SRTNet is a network entirely in time domain and does not depend on any Fourier transform.
Our main contributions are as follows: 
\begin{itemize}[leftmargin=18pt,itemsep=0.1pt,topsep=0.1pt,partopsep=0.1pt]
\item We innovatively apply the diffusion model as a stochastic refinement module for SE task and possess theoretical correctness. 
\item We introduce 
an ``enhance-and-refine" paradigm and use a joint network to implement it, which results in faster convergence and better speech quality. 
\item The proposed method eliminates the dependency that using STFT as the conditioner in previous methods\cite{A-study-SEAME,CD-SEAME} and achieves complete time-domain SE. Extensive experiments are carried out to show the effectiveness of the method.
\end{itemize}
\section{Proposed Methods}
\begin{figure}[tp]
 \centering
 \includegraphics[width=1\linewidth]{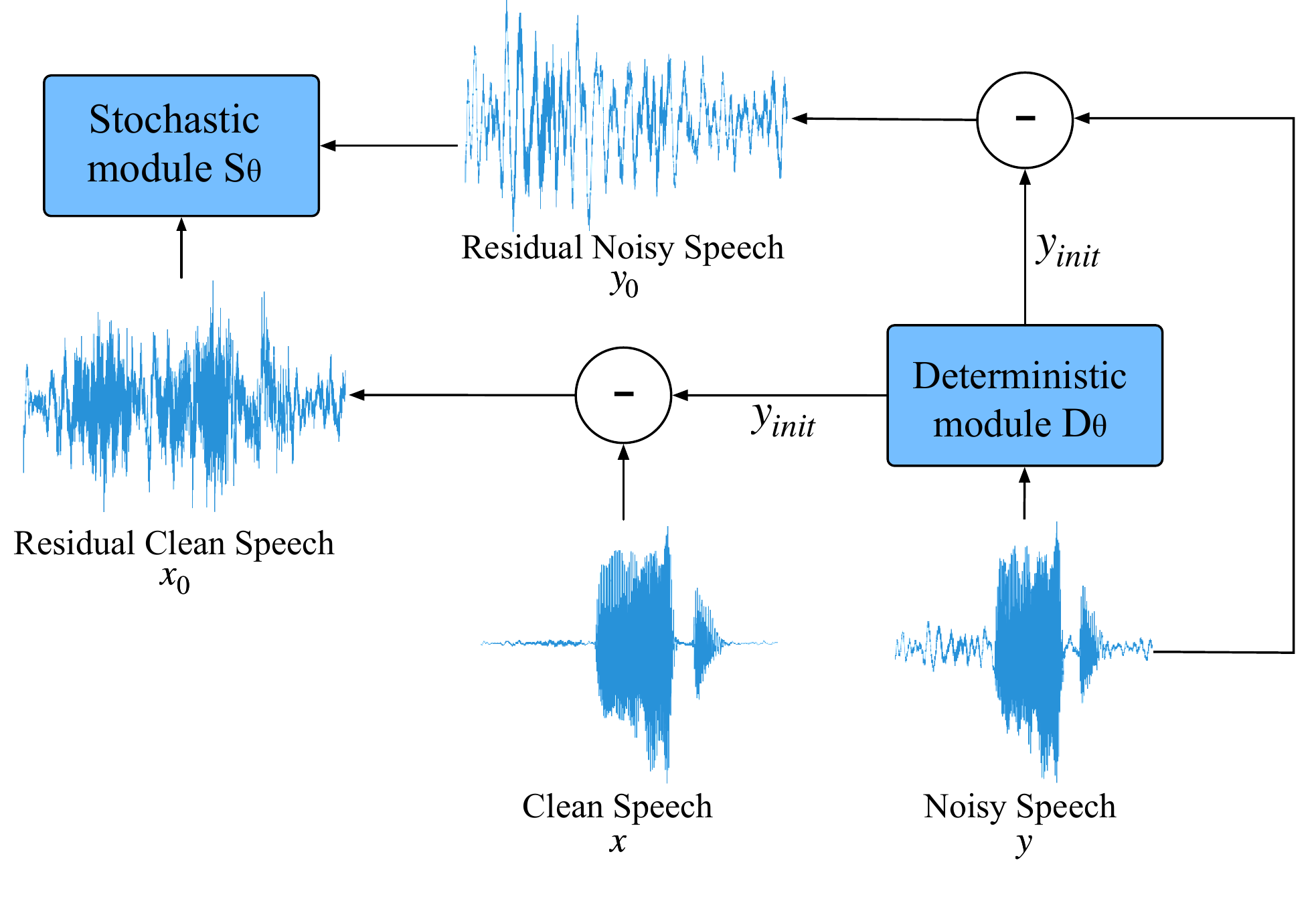} 
  \vspace{-0.3cm}
 \caption{Overall structure of SRTNet.}
 \label{fig:overall}
 \vspace{-0.3cm}
\end{figure}

Existing diffusion model based SE model generally trained directly on the original speech. Motivated by~\cite{Deblurring}, it is also reasonable for the SE task to learn the residual data. In the idea of ``enhance-and-refine", 
we add a deterministic module $D_\theta$ to the original conditional diffusion model. The block diagram of the overall structure is illustrated in Fig.~\ref{fig:overall}.
The noisy speech $y$ is initially enhanced by $D_\theta$ and we call its output $y_{init}$. Then the residual operations are enforced on the clean speech $x$ and the noisy speech $y$ with $y_{init}$ and the residuals $x_0, y_0$ are fed to stochastic module $S_\theta$. 
Because the output of the noisy speech passing through $D_\theta$ during sampling is deterministic when the parameters are determined, we call the $D_\theta$ deterministic module. Whereas in $S_\theta$, we need to sample from Gaussian distribution in diffusion model, which could produce different outputs, so we call the $S_\theta$ stochastic module.
\subsection{Diffusion and Reverse process of SRTNet}
\textbf{Diffusion.}
The diffusion process takes $y_0$ as a condition, and the noise in diffusion process contains not only Gaussian noise but also non-Gaussian noise in $y_0$ which is the residual of $y$ and $y_{init}$. The conditional diffusion process $q(x_t|x_0,y_0)$ is defined as follows:
\begin{align}
q(x_t|x_0,y_0)=\mathcal{N}(x_t;(1-m_t)\sqrt{\bar{\alpha}_t}x_0+m_t\sqrt{\bar{\alpha}_t}y_0,\delta_t I),
\label{eq:conditional diffusion process}
\end{align}
where $\Bar{\alpha}_t=\prod^t_{i=1}\alpha_i$, the noise schedule 
$\{\alpha_t\}^T_{t=1}$ is given. 
Additionally, $m_t$ is a interpolation ratio between 
the residual clean speech $x_0$ and the residual noisy speech $y_0$. The value of $m_t$ is defined as:
\begin{align}
m_t=\sqrt{(1-\bar{\alpha}_t)/\sqrt{\bar{\alpha}_t}}, 
\label{eq:original-value-of-m}
\end{align}
 where $m_0 = 0$ and $m_T \approx 1$. Therefore, the interpolation parameter $m_t$ gradually shifts the mean of Eq.~\eqref{eq:conditional diffusion process} from x-correlated to y-correlated with the diffusion process which satisfies a Markov chain, with details in~\cite{CD-SEAME}. And the variance ${\delta}_t$ is defined as:  
\begin{equation}
       \delta_t=(1-\bar{\alpha}_t)-m_t^2\bar{\alpha}_t. \label{eq:original-variance}
\end{equation}
\begin{algorithm}[tp]
  \caption{SRTNet Training.}
  \label{alg:training}
  \begin{algorithmic}
        \Repeat
        \State {Sample $({x}, {y}) {\sim}q_{\text{data}}, \epsilon{\sim}\mathcal{N}(0, \boldsymbol{I}), $ } 
        \State {$s{\sim}\text{Uniform}(\{1, \cdots, S\})$ , and $\sqrt{\bar{\alpha}}{\sim}\text{Uniform}(l_{s-1}, l_s)$}
        \State{Get ${{y}_{init}}$ through the deterministic module ${D_\theta}$, }
        \State{${{y}_{init}=D_\theta({y})}$}
        \State{Get two Residual: ${{x}_0={x}-{y}_{init}}$ and  ${{y}_0={y}-{y}_{init}}$}
        \State{Get ${{x}_t}$ according to Eq.~\eqref{eq:x_t}}
        \State {Take gradient descent step on}
        \State {$\nabla_\theta \parallel \frac{1}{\sqrt{1-\bar{\alpha}}}(m\sqrt{\bar{\alpha}}({{y_0}-{x_0}})+ \sqrt{\delta_t}\epsilon) - \epsilon_\theta({x}_t, {y}_0, \sqrt{\bar{\alpha}}) \parallel^2_2$}
        \Until converged
  \end{algorithmic}
\end{algorithm}
\noindent
\textbf{Reverse.}
In the reverse process, we start from $x_T$, with the condition $y_0$ and the variance $\delta_T$:
\begin{equation}
    p(x_T|y_0)=\mathcal{N}(x_T,\sqrt{\bar{\alpha}_t}y_0,\delta_T I).
\end{equation}
The reverse process also follows a Markov chain, so we can gradually obtain $x_0$ from $x_T$ by continuously executing the reverse process.
The parameterised conditional reverse process $p_\theta(x_{t-1}|x_t,y_0)$ is denoted as:
\begin{equation}
    \mathcal{N}(x_{t-1},c^x_t x_t+c^y_t y_0+c^\epsilon_t \epsilon_\theta(x_t,y_0,t),\tilde\delta_t I).
    \label{eq:Conditional reverse process}
\end{equation}
Note that the mean is parametrized as a linear combination of $x_t$, residual noisy speech $y_0$, and estimated noise $\epsilon_\theta$.
The coefficients $c^x_t, c^y_t $ and $ c^{\epsilon}_t$ are derived as follows:
 \begin{align}
    c^x_t&=\frac{1-m_t}{1-m_{t-1}}\frac{\delta_{t-1}}{\delta_t}\sqrt{\alpha}_t+(1-m_{t-1})\frac{\tilde\delta_t}{\delta_{t-1}}\frac{1}{\sqrt{\alpha}_t}\label{eq:c_xt}, \\
    c^y_t&=(m_{t-1}\delta_t-\frac{m_t(1-m_t)}{1-m_{t-1}}\alpha_t\delta_{t-1})\frac{\sqrt{\bar{\alpha}_{t-1}}}{\delta_t}\label{eq:c_yt}, \\
   c^{\epsilon}_t&=(1-m_{t-1})\frac{\tilde\delta_t}{\delta_{t-1}}\frac{\sqrt{1-\bar{\alpha}_t}}{\sqrt{\alpha_t}}\label{eq:c_et}.
 \end{align}
 The variance $\tilde{\delta}_t$ in Eq.~\eqref{eq:Conditional reverse process} can be derived from reverse diffusion process $p(x_{t-1}|x_t,x_0,y_0)$, and the detailed derivation is given in~\cite{CD-SEAME}:
\begin{equation}
    \tilde{\delta}_t=\delta_{t-1}-\Big(\frac{1-m_t}{1-m_{t-1}}\Big)^2\alpha_t\frac{\delta_{t-1}^2}{\delta_t}.
\end{equation}
\subsection{Training and Sampling of SRTNet}
\textbf{Training.}
The training process of the SRTNet is described in Algorithm~\ref{alg:training}. During the training phase, all parameters of diffusion process will depend on the noise level $\bar{\alpha}$ which is obtained by hierarchical sampling rather than time steps used in~\cite{CD-SEAME}. Specifically, a segment ${(l_{s-1},l_s)}$ is sampled from $s\sim U(\{1,\dots,S\})$ where $S$ is the length of 
the noise level schedule. Then the noise level $\bar{\alpha}$ is obtained from this segment by  sampling from the uniform distribution. The benefit of the diffusion model relying on the noise level in the training phase is that it allows us to sample with an arbitrary noise level schedule. The effectiveness of this approach is verified in~\cite{chen2020wavegrad}. Therefore, in  Eq.~\eqref{eq:conditional diffusion process}, the interpolation ratio $m$ is denoted as:
\begin{align}
    m=\sqrt{(1-\bar{\alpha})/\sqrt{\bar{\alpha}}},
\end{align}
and the variance ${\delta}$ is: 
\begin{equation}
       \delta=(1-\bar{\alpha})-m^2\bar{\alpha}. \label{eq:original-variance}
\end{equation}
Therefore, the expression of ${x}_t$ through reparameterizing the diffusion process Eq.~\eqref{eq:conditional diffusion process} can be denoted as:
\begin{equation}
 {x}_t =(1- {m})\sqrt{\bar{\alpha}}{x}_0 +  {m}\sqrt{\bar{\alpha}}{y}_0+\sqrt{\delta}\boldsymbol{\epsilon} 
\label{eq:x_t}.
\end{equation}
 To optimize the ELBO, we can directly model the mean of reverse process Eq.~\eqref{eq:Conditional reverse process}. But in practice, we generally model the noise added at a certain noise level  which is a unweighted variant of ELBO~\cite{DDPM}.
The objective function is defined as:
\begin{align}
    \mathbb{E} \parallel \epsilon_\ast
    - \epsilon_\theta(x_t, y_0,\sqrt{\bar{\alpha}}) \parallel^2_2,
 \label{eq:ELBO_optimized}
\end{align}
where $\epsilon_\ast$ is the noise from the combination of Gauss noise $\epsilon$ and the noise in $y_0$:
\begin{equation}
    \epsilon_\ast=\frac{m\sqrt{\bar{\alpha}}}{\sqrt{1-\bar{\alpha}}}{(y_0-x_0)}+ \frac{\sqrt{\delta}}{\sqrt{1-\bar{\alpha}}}\epsilon.
\end{equation}
\par
\noindent
\textbf{Sampling.}
The sampling process of the SRTNet is described in Algorithm~\ref{alg:sampling}. Through training, our model is able to efficiently model the noise at different noise levels. The noisy speech $y$ is first fed into the deterministic model $D_\theta$ to obtain $y_{init}$. Unlike the training process, here we obtain the noise level from the noise level schedule rather than hierarchical sampling. By iterating over the noise level schedule, we can run the reverse process through the reverse Markov chain. Afterwards, we obtain the residual clean speech ${x_0}$. The final enhanced speech will be obtained by summing ${x_0}$ and $y_{init}$.
\begin{algorithm}[tp]
  \caption{SRTNet Sampling.}
  \label{alg:sampling}
  \begin{algorithmic}
   \State{Get ${{y}_{init}}$ through the deterministic module ${D_\theta}$, }
    \State{${{y}_{init}=D_\theta({y})}$}
    \State{Get  Residual ${{y}_0={y}-{y}_{init}}$} 
  \State {Sample ${x}_T{\sim} \mathcal{N}({x}_T, \sqrt{\bar{\alpha}_T}{y}_0, \delta_T\boldsymbol{I}), $} 
  \For{$n=N, N-1, \cdots, 1$}
        \State {Compute $c^x_t, c^y_t$ and $c^{\epsilon}_t$ using Eq.~\eqref{eq:c_xt},~\eqref{eq:c_yt}, and~\eqref{eq:c_et}}
        \State {Sample $x_{t-1}\sim p_{\theta}(x_{t-1}|x_t, y_0)=$}
        \State {$\mathcal{N}(x_{t-1};c^x_tx_{t} + c^y_t y_0 - c^{\epsilon}_t \epsilon_\theta(x_t, y_0, \sqrt{\bar{\alpha}_t}), \tilde{\delta}_tI )$}
      \EndFor
      \State \textbf{return} ${x}_0 + {y}_{init}$
  \end{algorithmic}
\end{algorithm}
\subsection{Structure of deterministic and stochastic module}
The deterministic module and the stochastic module have a similar structure as in~\cite{A-study-SEAME,CD-SEAME}. Note that the conditioner and the noise level encoding here are only for the stochastic module. 
Moreover, we have made some other modifications to the structure.  Unlike~\cite{A-study-SEAME, CD-SEAME}, we have modified the structure with two main changes:\par
 Firstly, we directly use the waveform of the noisy speech $y$ as the conditioner instead of spectrogram. As a result, a complete time-domain SE is achieved. Moreover, it is also experimentally found that our model can converge faster, the detail in Sec.~\ref{Sec:Experiments}. \par
Secondly, we replace the time step encoding with a noise level encoding. But our encoding method differently from both in~\cite{kong2020diffwave, chen2020wavegrad}, the encoding method can be expressed by:
\begin{align}
  \sqrt{\bar{\alpha}}_{\text{encoding}}=&\Big[\sin{\Big(10^{\frac{0\times4}{63}}\sqrt{\bar{\alpha}}\Big)}, \dots, \sin{\Big(10^{\frac{63\times4}{63}}\sqrt{\bar{\alpha}}\Big)}, \nonumber\\
  &\cos{\Big(10^{\frac{0\times4}{63}}\sqrt{\bar{\alpha}}\Big)}, \dots, \cos{\Big(10^{\frac{63\times4}{63}}\sqrt{\bar{\alpha}}\Big)}\Big].
\label{eq:t_embedding}
\end{align}
Through this, we can obtain different encoding results with different noise level conditions as part of the input to the diffusion model.
 \begin{figure}[tp]
 \centering
 \includegraphics[width=1 \linewidth]{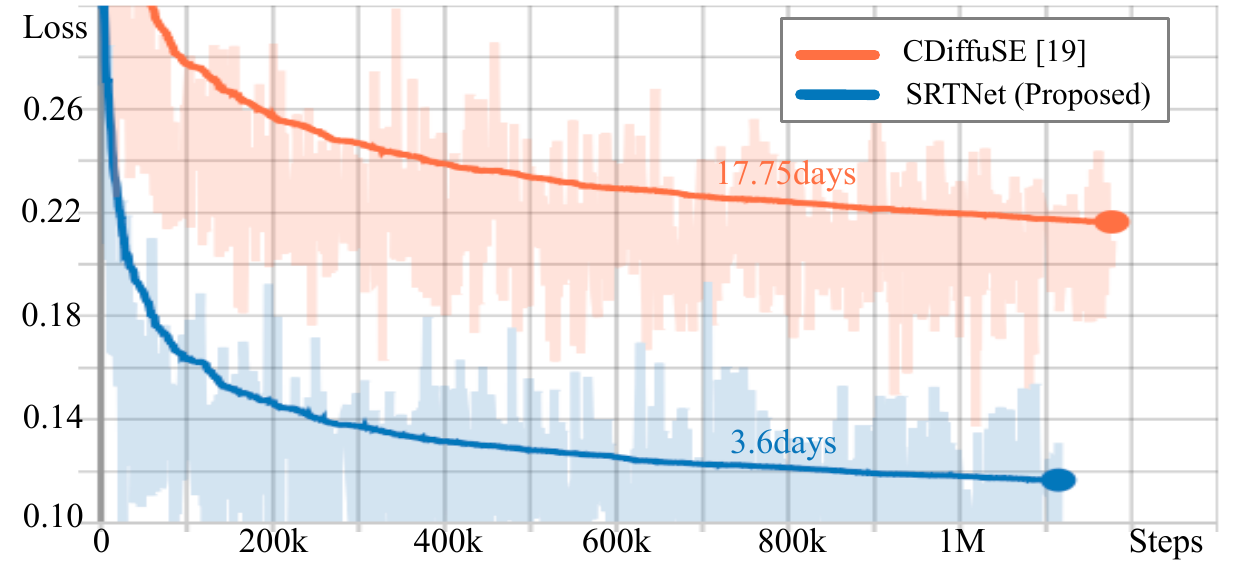} 
 \vspace{-0.3cm}
 \caption{Training curves of SRTNet and baseline~\cite{CD-SEAME} (Base). The numbers (17.75, 3.6) are the time  consumption to train the two models for 800k steps. }
 \label{fig:curve}\vspace{-0.3cm}
\end{figure}
\section{Experiments}
\label{Sec:Experiments}
\subsection{Experimental setup}
\label{subsec:Experimental Setup}
\subsubsection{Datasets}
The VoiceBank-DEMAND corpus~\cite{voicebank-corpus}, which includes 30 speakers from different accent regions in the UK and the US, is selected to evaluate the proposed method, with 28 speakers selected for training and 2 others for testing.
The training set consists of 11, 572 single channel speech samples, while the test set contains 824 utterances from 2 speakers (one male and one female). 
The signal-to-noise ratios (SNRs) of the training set are {0 dB, 5 dB, 10 dB, 15 dB}. The test set mixes with five unseen test noise types (all from the DEMAND database~\cite{DEMAND}) selected at {2.5 dB, 7.5 dB, 12.5 dB, 17.5 dB}. For the experiments, the original waveform is sub-sampled from 48 kHz to 16 kHz. 
WSJ0~\cite{wsj0} is another dataset commonly used for SE tasks.
The clean test set of WSJ0 contains 6 speakers (3 female and 3 male) with a total of 800 utterances. We use a clean test set of WSJ0 and 100 randomly selected noises from MUSAN~\cite{MUSAN} for mixing with ${- 5\sim 20}$ dB SNRs to verify generalization capability.
\vspace{-0.3cm}
\subsubsection{Performance metrics}
Four common SE evaluation metrics are used: perceptual evaluation of speech quality (PESQ)~\cite{pesq}, background intrusiveness (CBAK), prediction of the signal distortion (CSIG), and overall speech quality (COVL)~\cite{CBAK-CSIG-COVL}. The PESQ score ranges from -0.5 to 4.5, and the rest of the metrics range from 1 to 5. Higher score means better speech enhancement performance.
\vspace{-0.3cm}
\subsubsection{Training and sampling}
We train SRTNet 800k steps on two NVidia 3090 GPUs with Adam optimizer, learning rate is $2 \times 10 ^ {-4}$ and the batch size is set to 32. The inference noise level schedule is same to~\cite{CD-SEAME}. In order to recover the high frequency speech we combine the sampling results with the noisy speech with a ratio of 0.2 at the end of the reverse process~\cite{defossez2020real,abd2008speech}. 
Moreover, to avoid randomness, we infer the results several times and take the average value as the final result. For other models, we follow the training setups as in the original papers.
\subsection{Results}
\label{subsec:Results}
\subsubsection{Results from generative models in matched condition}
\label{subsec:Results on matched condition}
SRTNet and other recent generative models are trained on the VoiceBank-DEMAND dataset, and tested on the matched condition (training and testing on the same dataset). We can find that our method is the strongest generative model in Table~\ref{tab:SRTNet results}. Compared to other generative methods, we further narrow the gap with the regression-based discriminative models. In addition, compared with our baseline~\cite{CD-SEAME}, not only do we achieve better performance, but also greatly improve the convergence speed of the model. For example, when we train them 800k steps, the two models have almost converged. SRTNet consumes only one-fifth of the training time of~\cite{CD-SEAME} as shown in Fig.~\ref{fig:curve}. 
\vspace{-0.4cm}
 \begin{table}[H]
\footnotesize
\centering
\caption{SRTNet v.s. generative models (matched condition) }
\label{tab:SRTNet results}
\centering 
\begin{tabularx}{0.48\textwidth}{Xcccc}
    \toprule
    Method  &  PESQ($\uparrow$) & CSIG($\uparrow$)& CBAK($\uparrow$) & COVL($\uparrow$) \\
    \midrule	
    unprocessed  &  1.97 &  3.35 &  2.44 &  2.63\\
    \midrule
    SEGAN~\cite{SEGAN}  & 2.16 & 3.48 & 2.94 & 2.80 \\
    SERGAN~\cite{SERGAN}  &2.36 & 3.54 & 3.08 & 2.93\\
    SASEGAN~\cite{SAGAN} & 2.36 & 3.54 & 3.08 & 2.93 \\
    DSEGAN~\cite{phan2020improving} & 2.39 & 3.46 & 3.11 & 2.90 \\
    CP-GAN~\cite{CP-GAN} & 2.39 & 3.46 & 3.11 & 2.90 \\
    SE-Flow~\cite{Flow-Based1-SEAME} & 2.28 & 3.70 & 3.03 & 2.97 \\
    DiffuSE~\cite{A-study-SEAME} & 2.41 & 3.61 & 2.81 & 2.99   \\
    CDiffuSE~\cite{CD-SEAME}(Base) &2.44 & 3.66  & 2.83  &3.03\\
    CDiffuSE~\cite{CD-SEAME}(Large)  & 2.52 & 3.72  & 2.91 & 3.10\\
    \midrule
    SRTNet (Ours)   & \textbf{2.69} & \textbf{4.12} & \textbf{3.19} &\textbf{3.39}  \\
    \bottomrule
\end{tabularx}
\end{table}
\subsubsection{Results on generalizability to mismatched condition}
To verify that our model as a generative model has greater potential in generalizability than the discriminative model, we have done experiments on mismatched condition. The test set is obtained from a mixture of WSJ0 and MUSAN.
Table~\ref{tab:generalization_chime} shows the results. The performance of the discriminative models degrade greatly and our model has a significant advantage. The main reason is that the generative methods learn information about the data distribution rather than a mapping relationship learned by reducing a certain distance such as $L_p-loss$.
This feature allows SRTNet as a generative model to perform better.
\vspace{-0.4cm}
 \begin{table}[H]
\setlength{\tabcolsep}{3pt}
\footnotesize
\centering
\caption{SRTNet v.s. discriminative models (mismatched condition). The numbers in parentheses indicate the relative change in performance under mismatched condition and matched condition.}
\label{tab:generalization_chime}
\begin{tabularx}{0.48\textwidth}{Xcccc}
    \toprule
    Method & PESQ($\uparrow$) & CSIG($\uparrow$) & CBAK($\uparrow$) & COVL($\uparrow$) \\
    \midrule
    Unprocessed & 1.67 & 3.14 & 2.49 & 2.36 \\
    \midrule
    WaveCRN~\cite{hsieh2020wavecrn}& 1.89(-0.74) & 2.92(-1.03) & 2.51(-0.55) & 2.49(-0.80) \\
    Demucs~\cite{defossez2020real} & 1.76(-1.89) & 2.87(-1.12) & 2.56(-0.72) & 2.35(-0.97) \\
    Conv-TasNet~\cite{luo2019conv}& 2.08(-0.72) & 2.65(-0.03) & 2.26(-0.38) & 2.19(-0.32) \\
    \midrule
    SRTNet (Ours) & \textbf{2.56}(-0.11) & \textbf{3.91}(-0.21) & \textbf{2.85}(-0.34) & \textbf{3.17}(-0.22)  \\
    \bottomrule
\end{tabularx}
\end{table}
\subsubsection{Speech waveform and spectrogram analysis}
\label{subsec:Waveform and Speech Spectrogram Analysis}
\begin{figure}[tp]
 \centering
 \includegraphics[width=1\linewidth]{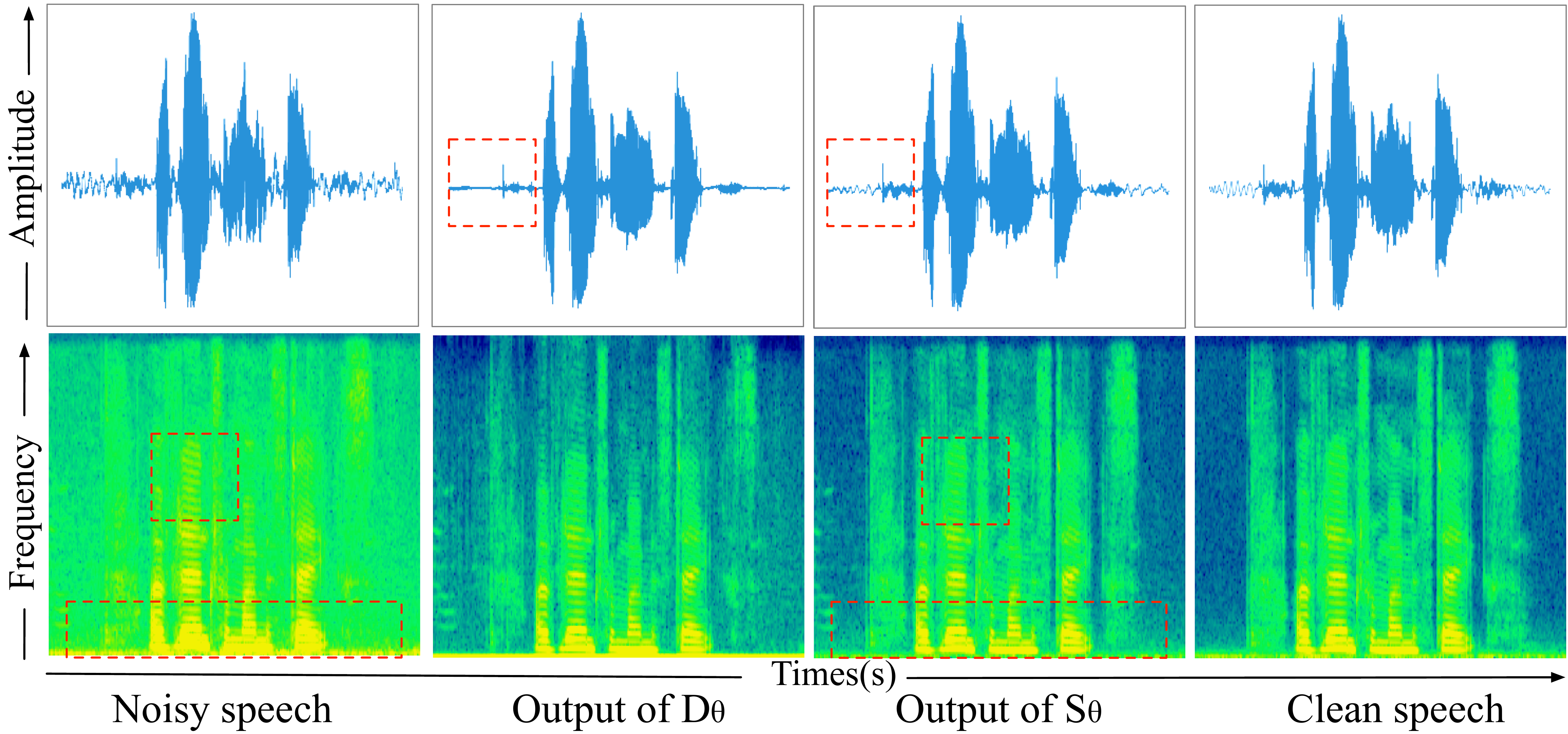}
 \caption{
 A test example (matched condition) enhanced by SRTNet and the output at different phases.
The output of each phase contains two parts: speech waveform and spectrogram.
}
 \label{fig:result}\vspace{-0.3cm}
\end{figure}

Fig.~\ref{fig:result} illustrates the waveforms and spectrograms of SRTNet output at different phases. By observing the waveforms, we can find that the noisy speech is more aggressively erased after the deterministic module, and then the information is refined by the stochastic module. From the two different phase of speech spectrograms, we can find that in the first phase most of the noise has been removed and our model can effectively deals with both high-frequency and low-frequency noise as shown in the red box. Although our model is a complete time domain approach, it performs well in the frequency domain, which is one of the strengths of our model.
\subsubsection{Ablation experiments}
 Here we carry out ablation tests to achieve deeper investigation into the contribution of each individual controlling factor.
When the best PESQ is obtained, we record the corresponding time consumption of training and sampling. Table~\ref{tab:Ablation experiments} demonstrates the experimental results.
The first experiment replaces the conditioner noisy waveform with noisy spectrogram as in~\cite{CD-SEAME}. The enhanced speech quality reduces from 2.69 to 2.59, indicating the effectiveness of the time-domain conditioning. We also see significant fast convergence and sampling speed in this setup. This is partly due to the absence of Fourier transform, and partly due to the time of up-sampling the speech spectrogram within the diffusion model. The second experiment replaces the noise level with time step compare to Eq.~\eqref{eq:Conditional reverse process} and the objective function expression is: 
\begin{align}
    \mathbb{E} \parallel \epsilon_\ast
    - \epsilon_\theta(x_t, y_0,t) \parallel^2_2.
 \label{eq:ELBO_optimized_a}
\end{align} \par
By the results we can find that using continuous noise level can effectively reduce the sampling time. The third experiment removes the deterministic module and we see a significant degradation of the enhanced speech quality, indicating the gain by ``enhance-and-refine".
 However, we also see the deterministic module increases the convergence time to some extent, because it adds an additional generation process. In summary, the deterministic module and time-domain conditioning contribute the most to the performance improvement.  
\label{subsec:Ablation experiments}
\vspace{-0.3cm}
\begin{table}[H]
\footnotesize
\centering
\caption{Ablation experiments. The PESQ and the relative convergence/Sampling time as metrics for our experiments. The original SRTNet convergence and sampling time to 1.0.}
\label{tab:Ablation experiments}
\vspace{0.1cm}
\begin{tabularx}{0.48\textwidth}{Xccc}
    \toprule
    Method & PESQ($\uparrow$) &Convergence/Sampling time($\downarrow$)  \\
    \midrule
    SRTNet  & 2.69 & 1.0 / 1.0 \\
    $-$waveform conditioner& 2.59 & 4.86 / 1.22\\
    $-$continuous noisy level &2.63& 1.34 / 3.66 \\
    $-$deterministic module & 2.58 & 0.83 / 0.88\\
    \bottomrule
\end{tabularx}
\end{table}
 \subsubsection{Variants of SRTNet}
\label{subsec:Variant experiments of SRTNet}
\begin{figure}[tp]
 \centering
 \includegraphics[width=1\linewidth]{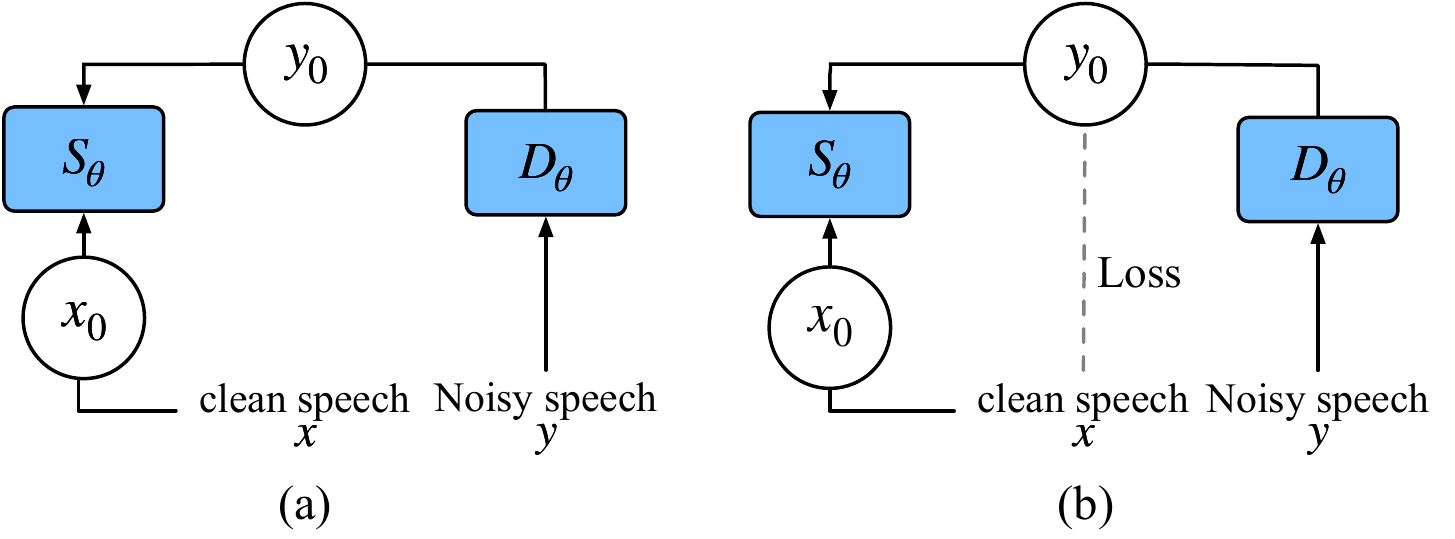}
 \vspace{-0.5cm}
 \caption{
  Two variants of SRTNet. (a) Residual-free SRTNet, (b) Residual-free SRTNet with an additional loss.
 }
 \label{fig:varies}
\end{figure}

Under the proposed ``enhance-and-refine" paradigm, SRTNet uses two residuals as input to the diffusion model. It is natural to directly use clean speech and the initially enhanced noisy speech as input. Therefore, we design a variant of SRTNet, namely, Residual-free SRTNet as show in Fig.~\ref{fig:varies}(a).
In this structure, it is intuitive to assume that the final enhanced speech depends heavily on the output of the deterministic module.
Therefore, another variant of the experiment is designed to ensure that the output of the deterministic model is as close to clean speech as possible by adding a loss between the output of the deterministic model and clean speech, as shown in Fig.~\ref{fig:varies}(b).
The results of the two variant experiments are shown in Table~\ref{tab:Variant_experiments}.
The experiments demonstrate that Residual-free SRTNet decreases in all metrics but still higher than other baselines.
This reflects the effectiveness of proposed ``enhance-and-refine" structure. When an additional loss function is added, the performance degrades dramatically. We attribute the reason to the additional loss causing the original generative model to be no longer pure and thus unable to learn the true data distribution.
In the previous models (SRTNet, Residual-free SRTNet), although there is not an individual loss function for the deterministic module, it still learns some useful information through the unified loss function.
\vspace{-0.3cm}
\begin{table}[H]
\setlength{\tabcolsep}{3pt}
\footnotesize
\centering
\caption{Variant experiments of SRTNet.}
\label{tab:Variant_experiments}
\vspace{0.1cm}
\begin{tabularx}{0.48\textwidth}{Xcccc}
    \toprule
    Method & PESQ($\uparrow$) & CSIG($\uparrow$) & CBAK($\uparrow$) & COVL($\uparrow$)\\
    \midrule
    Residual-free SRTNet &2.61&3.73&3.01&3.04\\
    Residual-free SRTNet$+$loss& 2.25 & 3.59 & 2.90 & 2.90\\
    SRTNet & \textbf{2.69} & \textbf{4.12} & \textbf{3.19} & \textbf{3.39}  \\
\bottomrule
\end{tabularx}\vspace{-0.3cm}
\end{table}
\section{Conclusion and Future Directions}
We propose SRTNet, a joint network for complete time domain speech enhancement. 
Our model achieves state-of-the-art performance for SE task in generative models while significantly reducing the time consumption of training. Furthermore, SRTNet has better generalization performance compared to those discriminative models.
In the next work we will use the proposed ``enhance-and-refine" paradigm to explore in the time-frequency domain, using the diffusion model to model the phase and the magnitude information separately.
However, our model also has some limitations, which we will investigate in our next work.  
When the noise situation is more complex and the signal-to-noise ratio is extremely low, our model may appear to be under-exerted, i.e., the initial enhancement of the noise interference in the deterministic module is not very thorough, resulting in the stochastic module that does not recover the clean speech well.
\label{sec:typestyle}
\begin{spacing}{0.85}
\bibliographystyle{IEEEtran}
\bibliography{refs}
\end{spacing}
\end{document}